\documentclass[conference, a4paper]{IEEEtran}
\IEEEoverridecommandlockouts

\ifCLASSINFOpdf
  \usepackage[pdftex]{graphicx}
  \DeclareGraphicsExtensions{.pdf,.jpeg,.png}
\else
\fi

\usepackage{cite}
\usepackage{amsmath,amssymb,amsfonts}
\usepackage{algorithmic}
\usepackage{graphicx}
\usepackage{textcomp}
\usepackage{url}
\usepackage[left=1.4cm, right=1.4cm, top=1.7cm]{geometry}
\usepackage{caption}
\usepackage{makecell}
\usepackage{booktabs}
\usepackage{anyfontsize}
\makeatletter
\newcommand*\titleheader[1]{\gdef\@titleheader{#1}}
\AtBeginDocument{%
\let\st@red@title\@title
\def\@title{%
\bgroup\normalfont\normalsize\centering\@titleheader\par\egroup
\vskip0.2em\st@red@title}
}
\makeatother

\makeatletter
\renewcommand{\fnum@figure}{Figure \thefigure}
\makeatother

\title{ {Like Uber or Like Buses? Economic Feasibility Analysis of UAM for Airport Access} \\
\vspace{0.5cm}
}

\titleheader{Second US-Europe Air Transportation Research and Development Symposium (ATRDS2026)}

\author{\IEEEauthorblockN{Shangqing Cao, Rishi Kumar Srinivasan, Raja Sengupta, Mark Hansen}
\IEEEauthorblockA{Department of Civil and Environmental Engineering \\
University of California, Berkeley \\
Berkeley, USA \\
caoalbert@berkeley.edu}
}

\renewcommand\IEEEkeywordsname{Keywords}
\IEEEaftertitletext{\vspace{-1\baselineskip}}

\begin{document}

\maketitle

\noindent \begin{abstract}

The airport access use case is a promising early-stage application for Urban Air Mobility (UAM). Understanding the operational paradigm of UAM at airports is crucial for making equitable and effective regulatory and management decisions. A central open question is whether UAM will be integrated into the airport transportation network as a conventional scheduled transit service, such as subways and rail, or as a Transportation Network Company (TNC) characterized by dynamic supply-demand matching. In this paper, we propose a two-stage framework for conducting an economic feasibility analysis of UAM networks. In the first stage, we introduce a joint-supply-demand variable pricing problem to evaluate the impact of dynamic pricing on UAM operations. This model uses a binary logit formulation to capture the trade-off between travel time advantages and fare levels. In the second stage, the determined demand is used as input for the Electric Urban Air Mobility Vehicle Routing Problem with Non-linear Charging Time (eUAMVRP-NL), which optimizes fleet scheduling and charging decisions to derive operating revenue and cost estimates. We apply this framework to a case study of the Los Angeles International Airport (LAX) access market with an eight-spoke vertiport network. Our results indicate that UAM operations benefit significantly from TNC-like management; a variable pricing policy can increase operating profits by more than 100\% compared to fixed-pricing schemes. Furthermore, we identify economies of stage length in longer UAM flights.\footnote{We open source the Python implementation of the joint-supply-demand pricing model, eUMAVRP-NL, and passenger arrival process model at \url{https://github.com/caoalbert/uam_system_model}.
}

\end{abstract}

\vspace{0.3cm}


\section{Introduction}

The airport access use case is one of the most promising early-stage markets for Urban Air Mobility (UAM). Studies and surveys have shown that passengers value predictable travel times for airport access trips and exhibit a higher willingness to pay for a travel time advantage in the airport access market \cite{coppola_urban_2024, hae_choi_exploring_2022}. Legacy carriers, including Delta, United, and American Airlines, have signed partnership agreements with various electric vertical takeoff and landing vehicle (eVToL) manufacturers to investigate the potential of UAM in the airport access market. We have reasons to believe the airport access market will become an early testing bed for UAM operations.

Mega-airports around the world function as multi-modal transportation hubs. Air transport at airports such as Paris Charles de Gaulle (CDG) and Hong Kong International Airport (HKG) is interwoven with rail, transit, and ferry services. The role that UAM plays in this web of transportation remains an open question. Will UAM operators behave similarly to transit, with a scheduled service and fixed pricing like subways and busses? Or will UAM operators function as transportation network companies (TNC), such as Uber, which specialize in real-time demand-supply matching powered by dynamic pricing? The paradigm of UAM operations can lead to significant implications for ground and air traffic management at airports and for equitable policy-making. 

In this paper, we seek to understand the paradigm of UAM operations through an economic feasibility study from the perspective of UAM operators. We discovered that UAM operations benefit from TNC-like management. A variable pricing policy that seeks to realize the value of the travel time advantage provided by UAM can more than double operating profit. We find that pricing schemes vary greatly across markets. Longer UAM flights lead to economies of stage length.

UAM operators, having ownership over both fleet management and pricing policies, are in a unique position to determine both the supply and demand in the marketplace. The first-stage problem of the two-step optimization framework proposed in this paper explores the profit-maximizing pricing scheme, considering passengers' sensitivity to cost and fleet scheduling policies that can accommodate the level of demand determined by the fare. The fare and level of demand obtained from the first-stage problem are then used as inputs to the second-stage Electric Urban Air Mobility Vehicle Routing Problem with Non-linear Charging Time (eUAMVRP-NL), through which we obtain detailed scheduling decisions for each aircraft in the fleet, taking into account non-linear battery charging time. We compute key operating metrics as well as implied infrastructure requirements using solutions to eUAMVRP-NL, all of which are used to assess the economic feasibility of a UAM service for the LAX airport access market. 

The rest of the paper is structured as follows: Section \ref{sec:prior_work} introduces existing work on estimating the economic feasibility, pricing, and the fleet scheduling problem for UAM operations. Section \ref{sec:method} introduces our proposed framework for estimating the economic feasibility of a UAM service and the various optimization problems we use in the process. Section \ref{sec:case_study} describes the case study network, the exogenous parameters for the optimization model, and the cost metrics. We discuss the results and insights into UAM operations in Sec. \ref{sec:results}.

\section{Prior Work}
\label{sec:prior_work}

\subsection{Economic feasibility of UAM}

The economic feasibility of various UAM operations has been studied with varying levels of scope and depth to review different operational models. Choi et al. use a multinomial logit model (MNL) on Incheon ground access data to identify the additional value to be captured with time saved for commuters in an airport shuttle use-case, which permits a \$57 premium over ground transport for 50 minutes of reduced travel time \cite{hae_choi_exploring_2022}. This places the maximum fare in Seoul at about 96 USD ($\sim$\$3.30/mile) to 108 USD based on a 30-40 min reduction in travel time. In addition, the study proposes a \$60 fare per seat to achieve an internal rate of return of 2.25\% over 20 years. Lv et al. achieve daily operational profitability in Beijing through a mixed-integer linear programming (MILP) revenue maximization at a fixed fare of 2.62 RMB/km \cite{lv_urban_2024}. It captures 90.7\% to 94.8\% of the expected orders. Earlier airport use-case studies, such as Uber Elevate's 2016 report \cite{holden_fast-forwarding_2016}, suggested USD 3.60 - USD 10.90 /pax-mile with a load factor of 50\% and a revenue flight ratio of 1:6. However, there was no indication of long term economic feasibility at that price level. While the estimates can be widely different, there currently exists a gap in the literature for systematically understanding the feasibility of a UAM service considering both pricing, which determines revenue, and fleet scheduling, which determines the operating cost.

\subsection{Joint Supply-Demand Optimization}

The earlier UAM studies do not address the need for a combined optimization of supply and demand, an ability observed only in air-taxi operations and autonomous vehicles. The operator has the ability to control both the inputs of the MNL, including travel time and price, along with supply considerations such as fleet size and repositioning. Wen et al. propose relating granular and dynamic demand data to autonomous mobility dispatching \cite{wen_value_2019}. The simulation suggests that the ability to re-balance the fleet dynamically based on demand can increase revenue. Li et al. used a graph neural network reinforcement learning framework with bi-level control and achieved 10\%-25\% higher profitability and customer service compared with the equivalent under non-joint optimization \cite{li_learning_2025}. While a comparison with a non RL based controller is not provided, the results validate the endogenous approach to UAM optimization when given the ability.

\subsection{UAM Fleet Scheduling Problem}

The UAM fleet scheduling problem studies the optimal routing of a fleet of UAM aircraft to assist in making strategic decisions in network design, fleet sizing, and infrastructure sizing. Insights from the solution to the fleet scheduling problem can be used to make financial forecasts and estimate the economic feasibility of a given UAM service. Lv et al. formulates a MILP that 
solves a joint problem of UAM network design and fleet scheduling \cite{lv_urban_2024}. The objective of the model is to maximize operating profit by choosing vertiport locations for a given UAM demand while simultaneously designing flight schedules connecting the selected vertiports, considering repositioning flights. Gottschalg et al. propose a two-stage framework in which the first-stage problem solves for strategic decisions such as fleet sizing, vertiport location, and vertiport sizing, assuming probabilistic demand profiles. The second-stage problem then solves for aircraft scheduling policies \cite{gottschalg_vertiport_2026}. The formulation includes a Value-at-Risk (VaR) constraint on profits in the first-stage problem to ensure the placement of vertiports can lead to profitable operations. However, pricing is treated as an external parameter rather than a decision variable. Husemann et al. investigates optimal battery weight, charger power, and fleet mix through a charging scheduling optimization model that takes MATSim-derived flight schedules as input and outputs the most cost-effective battery charging policy \cite{husemann_analysis_2024}. The flight schedules are external to the optimization model, and the scheduling of flights, including repositioning flights, is generated based on heuristics in MATSim. Kotwicz et al. adopt a bi-level optimization framework to first determine the minimal fleet size and the optimized flight schedules and then identify the minimum requirement for vertiport sizing \cite{kotwicz_herniczek_fleet_2024}. Roy et al. employ a multi-commodity network flow model to maximize the total number of flights that can be served by a given fleet \cite{roy_flight_2022}. 

\section{Methodology}
\label{sec:method}

Our framework consists of two main modules. In the first stage, we leverage a joint-supply-demand pricing problem to compute the profit-maximizing fare of UAM flights and the level of demand while ensuring that the demand can be sufficiently satisfied by a given fleet of vehicles. From the solution to the joint-supply-demand profit maximization problem, we recover the hourly passenger demand as well as the determined fare for UAM flights within each discrete time period. We then use the determined passenger arrival rate and the fare to generate passenger arrival processes using the demand generation model proposed in \cite{cao_fleet_2024}. In the second stage, we compute the optimal routing of the fleet of aircraft to serve the revenue flights using a vehicle routing formulation of the UAM fleet scheduling problem, eUAMVRP-NL, to compute a realistic and feasible scheduling policy for the vehicles. We recover key operational metrics from eUAMVRP-NL, including but not limited to revenue, the number of passengers served, load factor, aircraft hours, aircraft miles, the percentage of repositioning flights, and implied infrastructure size. These metrics are used to calculate both the capital and operating costs of the UAM service, with which we can estimate the economic feasibility of the airport access use case for UAM.

\subsection{Joint-Supply-Demand Pricing Problem}

The goal of the joint-supply-demand pricing problem is to determine the profit-maximizing fare for UAM flights. As there exists a one-to-one mapping between fare and market share, as assumed in Discrete Choice Models (DCM), pricing is equivalent to determining the market share of UAM and subsequently passenger demand. We consider the coupled decision between pricing and fleet scheduling, as the operators need to ensure that available aircraft are allocated to profitable markets when needed. 

Let $\mathcal{I}$ be a set of spatial-temporal origin-destination (OD) pairs, with each element being a tuple of origin vertiport, destination vertiport, and time of departure. Let $\mathcal{J}$ be a set of transportation mode alternatives. We assume a scenario in which UAM is competing with TNCs. Therefore, $|J|$ = 2. We assume the total addressable market for UAM is known; therefore, we are not considering induced demand. We use $d_{i}$ to denote the total addressable market, in number of passengers, for $i\in \mathcal{I}$.  Passengers have to choose between TNC and UAM. The objective function of the joint-supply-demand pricing problem can be written as:

\begin{equation}
    \underset{\theta_{i}^{\text{UAM}}, r_{i}^{\text{UAM}}, x_{i}, y_{ij}}{\text{max}} \sum_{i \in \mathcal{I}} \sum_{j \in \mathcal{I} \setminus i} r_{i}^{\text{UAM}} d_{i} \theta_{i}^{\text{UAM}} - c_{i}x_{i} - p_{ij} y_{ij}
\label{eq:1_obj_pricing}
\end{equation}

where $r_{i}^{\text{UAM}}$, $\theta_{i}^{\text{UAM}}$ are decision variables of the UAM fare and market share in OD, $i\in \mathcal{I}$, respectively. $x_{i}$ is the number of aircraft scheduled to serve passengers on $i\in \mathcal{I}$, and $c_{i}$ is the operating cost of a flight on $i \in \mathcal{I}$. $y_{ij}$ is the number of aircraft to reposition to the origin of $j \in \mathcal{I}\setminus i$ from the destination of $i\in \mathcal{I}$ after completing $i$. The cost of a repositioning flight from the destination of $i$ to the origin of $j$ is $p_{ij}$. The objective function is the sum of the differences between revenue and operating costs, which considers the cost of operating both revenue and repositioning flights. 

We use a binary logit model to estimate the market share of UAM as a function of travel time and fare. We express the utility of a transportation mode $j\in \mathcal{J}$ on $i\in \mathcal{I}$ as  

\begin{equation}
    V^{j}_{i} = -t_{i}^{j} - \frac{1}{p_{i}} r_{i}^{j}
\end{equation}

where $t_{i}^{j}$ is the travel time of mode $j$ in spatial-temporal OD $i$ and $p_{i}$ is the value of time of passengers on OD $i\in \mathcal{I}$.

By the Independence of Irrelevant Alternatives (IIA) property of the multinomial logit (MNL) model, which implies that the relative choice probability between any two alternatives depends only on their respective utilities, we can express the relationship between the utilities of UAM and TNC as follows:

\begin{equation}
\label{eq:3_sub}
        V_{i}^{\text{UAM}} = \ln \theta_{i}^{\text{UAM}} - \ln \theta_{i}^{\text{TNC}} + 
    V_{i}^{\text{TNC}}
\end{equation}



We can expand (\ref{eq:3_sub}) into 

\begin{equation}
       -t_{i}^{\text{UAM}}-\frac{1}{p_{i}}{r_{i}^\text{UAM}} = \ln \theta_{i}^{\text{UAM}} - \ln\theta_{i}^{\text{TNC}} + 
    V_{i}^{\text{TNC}}
\end{equation}

and by the assumption that $\theta_{i}^{\text{UAM}} + \theta_{i}^{\text{TNC}} = 1$

\begin{equation}
\label{eq:5_ri}
       {r_{i}^\text{UAM}} = -p_{i} [\ln \theta_{i}^{\text{UAM}} - \ln(1-\theta_{i}^{\text{UAM}}) + 
    V_{i}^{\text{TNC}} + t_{i}^{\text{UAM}}]
\end{equation}

Therefore, we can substitute the decision variable $r_{i}^{\text{UAM}}$ in (\ref{eq:1_obj_pricing}) with $\theta_{i}^{\text{UAM}}$ using (\ref{eq:5_ri}). The substitution converts the domain of the problem from pricing to determining the market share of UAM. This is a technique discussed in detail in \cite{kim_strategic_2025, cao_joint_2026}. 

Waiting time at vertiports makes up a significant portion of the UAM travel time, depending on the levels of service (LoS) provided. Therefore, we model $t_{i}^{\text{UAM}}$ as a decision variable that depends on the number of flights scheduled. We consider the impact of LoS on travel time, assuming that people arrive at the vertiport within each time interval $T$ uniformly. Therefore, the average travel time can be expressed as:

\begin{equation}
\label{eq:ti}
    t_{i}^{\text{UAM}} = a_{i}^{\text{UAM}} + \frac{T}{x_{i}} \times \frac{1}{2}
\end{equation}

where $a_{i}$ is the total UAM travel time on $i\in \mathcal{I}$ that excludes waiting time. The final objective function can then be written as:

\begin{multline}
    \underset{\theta_{i}^{\text{UAM}}, r_{i}^{\text{UAM}}, x_{i}, y_{ij}}{\text{min}} \sum_{i \in \mathcal{I}} \sum_{j \in \mathcal{I}\setminus i} d_{i}p_{i}\theta_{i}^{\text{UAM}} (V^{\text{TNC}}_{i} + a^{\text{UAM}}_{i} +   \ln{\theta_{i}^{\text{UAM}}} \\ - \ln{(1-\theta_{i}^{\text{UAM}}})) + d_{i}p_{i}\theta_{i}^{\text{UAM}} \frac{T}{2x_{i}} + x_{i}c_{i} + y_{ij}p_{ij}
\end{multline}

To model the coupling relationship between passenger demand and fleet scheduling, we define the joint-supply-demand pricing problem using the network formulation of the classic assignment problem. Let $G(W,E)$ be a time space network. We define the set of nodes W as:

\begin{equation}
    W = \{i_{start} |i\in \mathcal{I}\} \cup \{i_{end} | i\in \mathcal{I}\} \cup \{source, sink\}
\end{equation}

and the set of edges is as follows:

\begin{multline}
        E = \{(i_{start}, i_{end}) | i\in \mathcal{I}\} \cup \{(source, i_{start}) | i\in \mathcal{I}\} \cup \\ 
        \{(i_{end}, sink) | i \in \mathcal{I}\} \cup \{(i,j) | \Tilde{a}_{i} + \Tilde{f}_{ij} \leq \Tilde{d}_{j} \ i \in \mathcal{I}, j \in \mathcal{I} \setminus i\}
\end{multline}

where $\Tilde{a}_{i}$ is the scheduled arrival time of the flight on $i\in \mathcal{I}$, and $\Tilde{d}_{j}$ is the scheduled departure time of the flight on $j \in \mathcal{I}\setminus i$. We use $\Tilde{f}_{ij}$ to represent the repositioning flight time between the destination of $i\in \mathcal{I}$ and the origin of $j\in \mathcal{I}\setminus i$. In other words, we construct a repositioning edge between any two spatial-temporal OD nodes in which there exists sufficient time for a repositioning flight. $\Tilde{f}_{ij}=0$ if the destination of $i$ is the same as the origin of $j$. Decision variable $x_{i}$ is equivalent to the flow $f_{ij}$ on edge $(i,j) \in \{(i_{start}, i_{end}) | i\in \mathcal{I}\}$, and $y_{ij}$ is equivalent to the flow on the repositioning edges $ \{(i,j) | \Tilde{a}_{i} + \Tilde{f}_{ij} \leq \Tilde{d}_{j} \ i \in \mathcal{I}, j \in \mathcal{I}\}$.

The optimization problem is subject to the following constraints:

\begin{equation}
\label{cons:demand}
        N x_{i} \geq d_{i}\theta_{i}^{\text{UAM}} \quad \forall i\in \mathcal{I}
\end{equation}

\begin{equation}
\label{cons:non_negative_flow}
        f_{ij} \geq 0 \quad  \forall (i,j) \in E 
\end{equation}

\begin{equation}
\label{cons:flow_conservation}
        \sum_{j} f_{ji} - \sum_{j} f_{ij} = n_{i} \quad \forall i \in W
\end{equation}

Constraint (\ref{cons:demand}) ensures that the number of seats provided in the spatial-temporal OD market $i\in \mathcal{I}$ must be greater than or equal to the number of passengers captured by the fare chosen. $N=4$ is the assumed seat capacity of the eVToL aircraft. Constraint (\ref{cons:non_negative_flow}) specifies that the flow in network $G$ must be non-negative. Constraint (\ref{cons:flow_conservation}) guaranties the conservation of flow, and $n_{i}$ is the node capacity of $i\in W$. The sink and source nodes have a node capacity of $-K$ and $K$, respectively, with $K$ being the fleet size. The node capacity of other nodes is set to 0.

\subsection{Passenger Arrival Process Generation}

We recover the fare and the market share of UAM for each spatial-temporal OD pair from solutions to the joint-supply-demand pricing optimization model. To model realistic operation, we adopt the autoregressive passenger arrival process model proposed in Cao et al. \cite{cao_fleet_2024} to obtain the arrival time of passengers at each vertiport. The autoregressive Poisson process takes the expected rate of arrival, determined by the joint-supply-demand pricing optimization model, to generate the arrival times of each passenger. We then apply the deterministic dispatch rule described in \cite{cao_fleet_2024} to create revenue UAM flights given the passenger arrival sequence. By the end of this procedure, we obtain a set of revenue flights as well as the number of passengers and the underlying revenue associated with the flight. We use the Electric Urban Air Mobility Vehicle Routing Problem with Non-linear Charging Time (eUAMVRP-NL) proposed in \cite{cao_urban_2026} to compute the charge and flight schedules of each aircraft in the fleet.

\subsection{Electric Urban Air Mobility Vehicle Routing Problem with Non-linear Charging Time (eUAMVRP-NL)}

For a given set of revenue flights, $\mathcal{I}$, and a fixed fleet size of $K$, the goal of eUAMVRP-NL is to find a flight schedule that includes repositioning flights and a charging schedule for which certain operational objectives, such as revenue, are maximized. Each revenue flight $i\in \mathcal{I}$ is defined by its origin vertiport, destination vertiport, flight duration, energy consumption, and scheduled departure time. The problem is defined in a directed time-space graph where each element of the demand set $\mathcal{I}$ is a node. Coupled with considerations for charging, repositioning, and the scheduled departure time of UAM flights, the objective is to find an optimal path for each aircraft on the time-space graph from the source to the sink node, during which the demand nodes are visited.

We assume a homogeneous fleet, and each aircraft is equipped with the same battery. We also assume a homogeneous set of chargers installed at the vertiports. We do not consider infrastructure constraints such as the number of landing pads and parking pads. Our model is able to provide the decision maker with estimates of infrastructure requirements. We impute the implied infrastructure requirement for the number of chargers and TLOFs from the solution to the optimization problem, an approach that is common in the existing literature \cite{shon_optimization_2025, cao_fleet_2024}.

\begin{figure}[h!]
    \centering
    \includegraphics[width=1\linewidth]{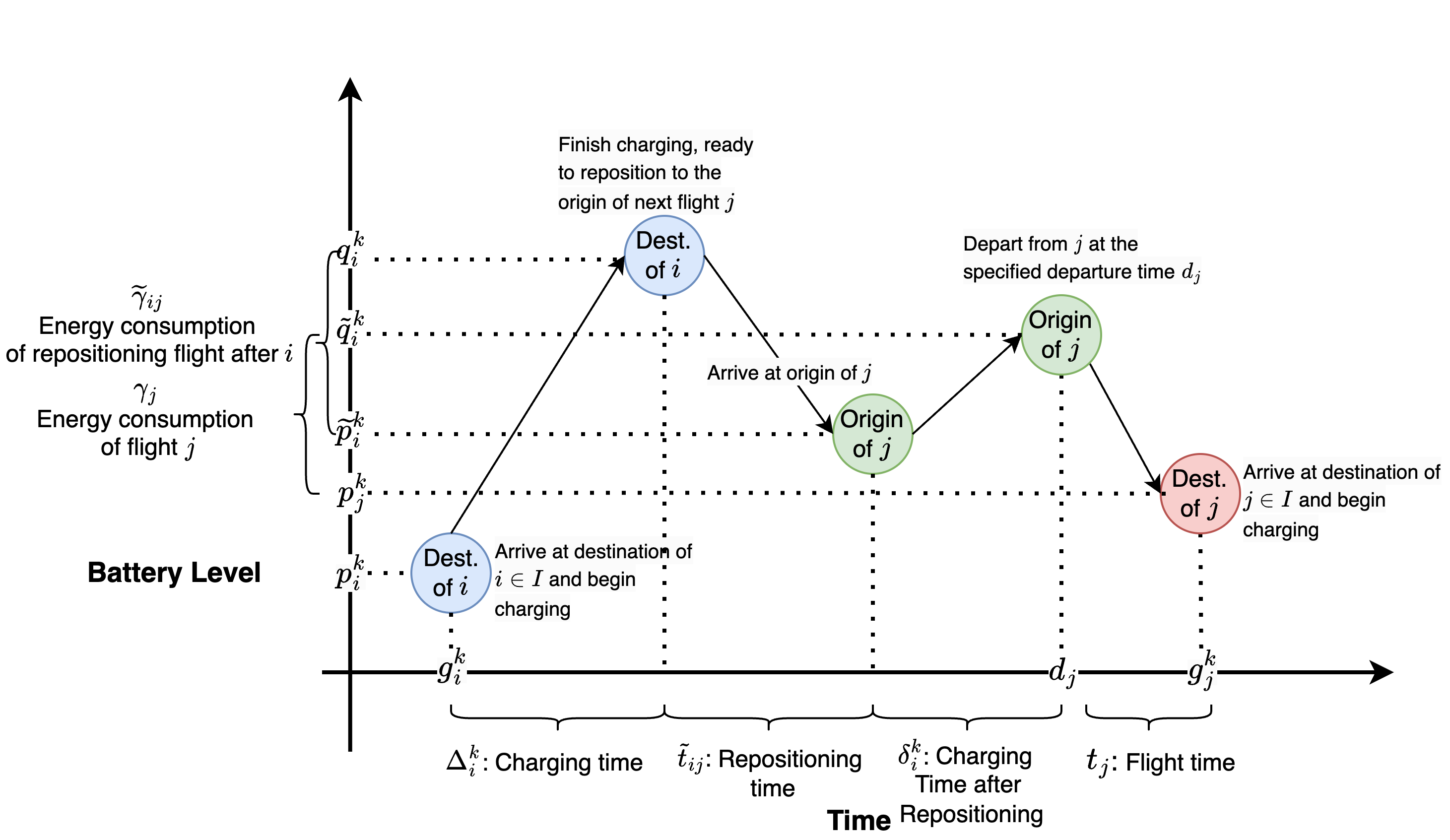}
    \caption{Cycle of an aircraft between completing a flight $i$ and completing the subsequent flight $j$. In between, aircraft can be charged for a maximum number of two times, reposition if needed, and serve the upcoming revenue flight. Nodes highlighted in same color denote the same vertiport.}
    \label{fig:vrp}
\end{figure}

Figure \ref{fig:vrp} illustrates the complete transition process between when a vehicle completes the previous revenue flight and before it serves the next revenue flight. Each aircraft has two opportunities to be charged: one before and one after the repositioning flight. This is important due to the non-linearity in the battery charging model. Operating the aircraft in a lower State of Charge (SoC) leads to time savings in charging \cite{montoya_electric_2017}. The marginal SoC gained per unit of time decreases as SoC increases. 

We adopt a three-index vehicle routing formulation with time-window constraints that track the time horizon of each vehicle in the fleet and the SoC level. The problem is solved using a Cluster-First Route-Second heuristic that decomposes the K-vehicle VRPs problem into k-numbers of 1-vehicle VRPs. Further details about the eUAMVRP-NL model can be found in Cao et al. \cite{cao_urban_2026}.

\section{LAX Case Study}
\label{sec:case_study}

\subsection{UAM Network and Passenger Demand}

\begin{figure}[h!]
    \centering
    \includegraphics[width=1\linewidth]{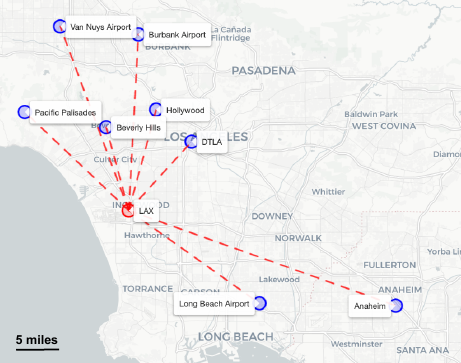}
    \caption{Area of study. The hub vertiport is located at LAX while the spokes vertiports are distributed in the Greater Los Angeles metropolitan region.}
    \label{fig:area_of_study}
\end{figure}

As shown in Fig. \ref{fig:area_of_study}, we consider an eight-spoke star network centered on LAX as the case study market. The selection of spoke vertiports is based on the top ten most common passenger origin cities reported by the 2019 LAWA Passenger Survey \cite{unison_consulting_2019_2019}.

\begin{figure}[h!]
    \centering
    \includegraphics[width=1\linewidth]{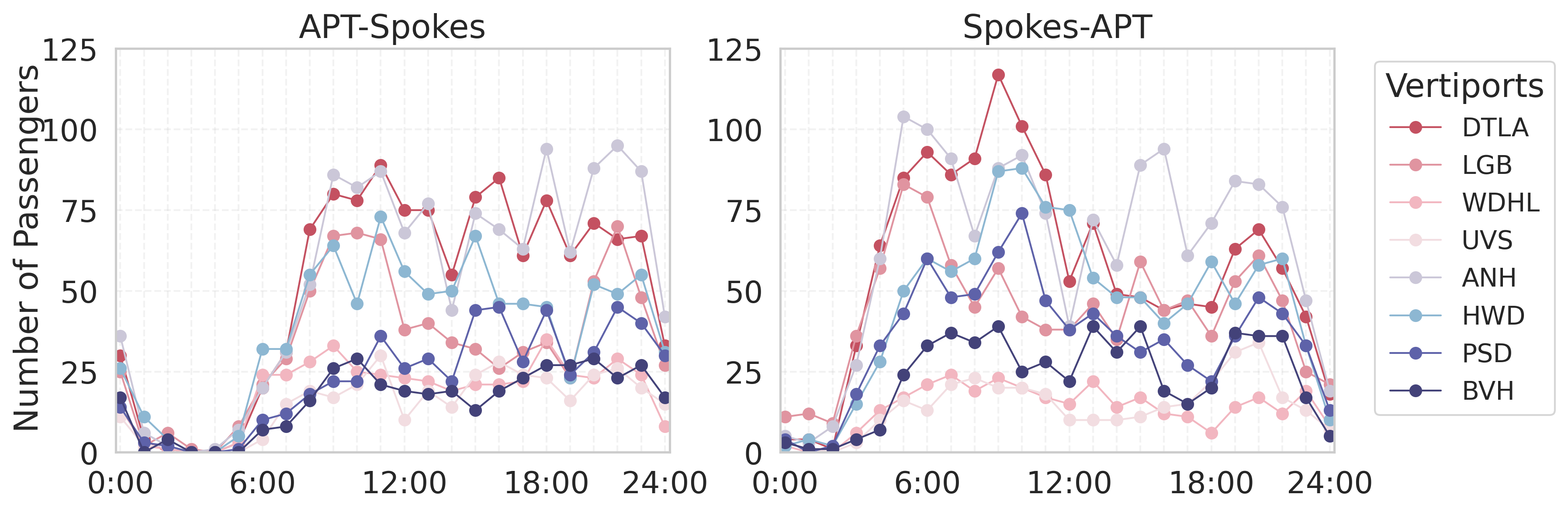}
    \caption{Total addressable market by time of the day across OD markets.}
    \label{fig:address}
\end{figure}

Based on this configuration, we estimate the total addressable market for the proposed UAM system to be approximately 4.78 million passengers annually. In 2019, LAX handled 77.5 million origin–destination (OD) passengers \cite{bureau_of_transportation_statistics_t-100_2019}. According to the Ground Transportation Monthly Report published by Los Angeles World Airports (LAWA), 33.33\% of OD passengers used transportation network company (TNC) services for airport access trips during that year \cite{lawa_angeles_2019}. The 2019 LAWA Passenger Survey further identifies the ten most common passenger origin cities and the corresponding share of travelers accessing the airport from each location \cite{lawa_angeles_2019}. On a typical day of the year, based on the median number of departure and arrival seats offered by legacy airlines at LAX. Visualized in Fig. \ref{fig:address}, the median daily addressable market is estimated to be 13,043 passengers across the two directions to and from LAX.

\subsection{UAM and TNC Travel Time}

As explained previously in (\ref{eq:ti}), the total UAM trip time is the sum of a constant term $a_{i}^{\text{UAM}}$ and a waiting time that is determined by the flight schedules. We assume $a_{i}^{\text{UAM}}$ to be:
\begin{multline}
        a_{i}^{\text{UAM}} = \text{Flight Time}_{i} + \text{Transition Time} + \\ \text{Last or First Mile Trip Time}_{i}
\end{multline}

where the transition time is assumed to be 10 minutes, including the passenger egress and ingress time at vertiports. A UAM trip is multi-modal. Passengers need to reach their final destinations or depart for spoke vertiports using a ground transportation mode. We randomly sample 10 origins (destinations) within a radius of 4 miles of each spoke vertiport. We use the Google Maps API to obtain the ground travel time between each sampled point and the vertiport. We compute the average hourly first and last mile travel time. Similarly, we use the same sampled points to compute the average hourly ground travel time to and from LAX. This becomes the TNC travel time that we use in the binary logit model for estimating the market share for UAM. UAM flight time is computed using the state-of-the-art UAM network simulator, \textit{Vertisim} \cite{onat_evaluating_2024}, which relies on the kinematic equations described in \cite{sripad_promise_2021}.

\begin{figure}[h!]
    \centering
    \includegraphics[width=1.05\linewidth]{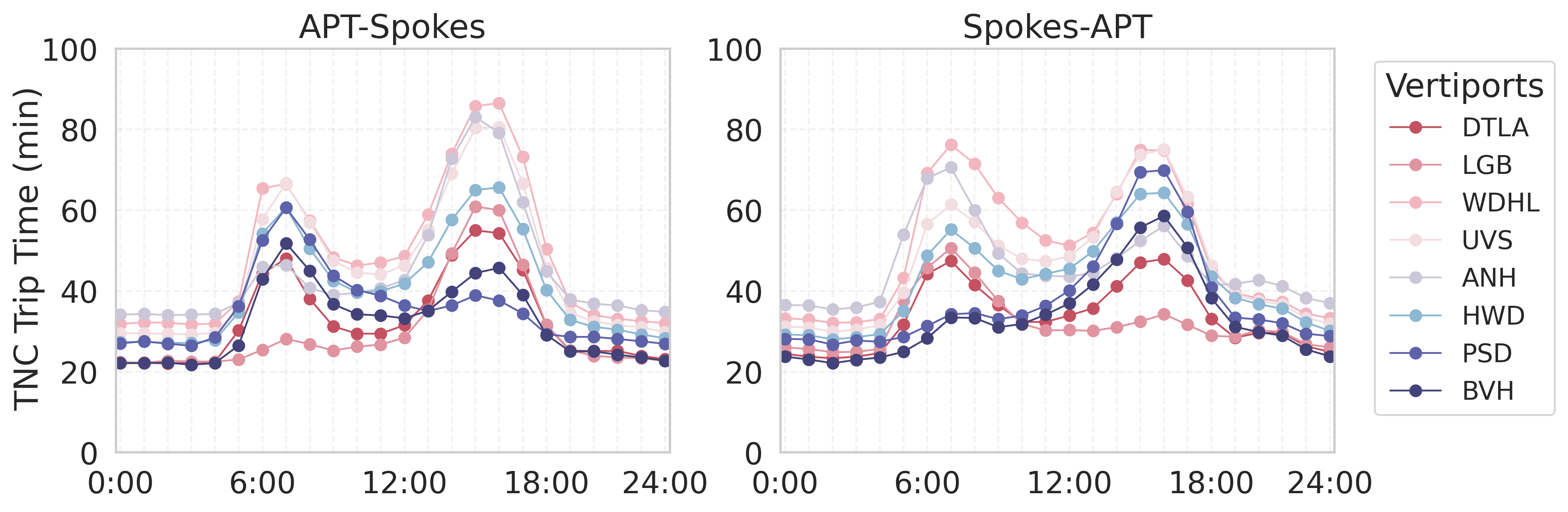}
    \caption{TNC trip time by hour.}
    \label{fig:tnc_trip_time}
\end{figure}

\begin{figure}[h!]
    \centering
    \includegraphics[width=1.05\linewidth]{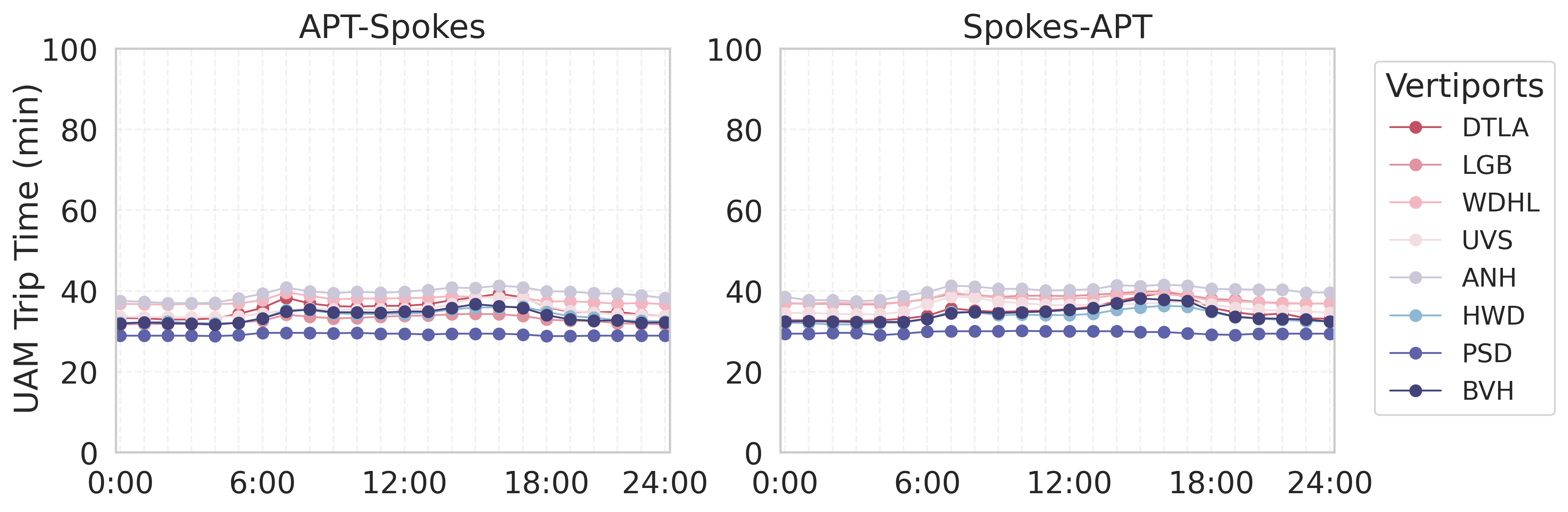}
    \caption{UAM trip time by hour. This includes an assumed 5 minute waiting time at vertiport.}
    \label{fig:uam_trip_time}
\end{figure}

Figure \ref{fig:tnc_trip_time} and \ref{fig:uam_trip_time} show the hourly travel time using UAM and TNCs by OD pairs during different times of the day. We can observe a clear increase in ground travel time during the morning and afternoon, caused by congestion. In comparison, UAM trip times exhibit less variation across different hours of the day, though minor differences still exist due to first and last mile trips, which occur on the ground. We see that although UAM offers a travel time advantage, it is a fragile one. In the early mornings and late at night, travel time using UAM is comparable to that of TNC. Therefore, finding the optimal pricing to capitalize on the travel time advantage is important for driving the profitability of the service.

\vspace{-0.4cm}

\subsection{TNC Fare}

We obtained the TNC fare for airport access trips using Uber's price estimator tool, reported below in Tab \ref{tab:uber}.

\begin{table}[h]
    \centering
    \caption{Average Uber fare for the 8 OD markets considered in the proposed network. Air distance is the Haversine distance between LAX and the spoke vertiport of the given market. Revenue per ground mile (RpGM) is defined as the ratio between Uber fare and the ground distance. Revenue per air mile (RpAM) is defined as the ratio between Uber fare and the air distance.}
    \label{tab:uber}
    \begin{tabular}{l r r r r r}
        \toprule \textbf{OD Market}
        & \textbf{Ground} & \textbf{Air} & \textbf{Average} & \textbf{RpGM} & \textbf{RpAM} \\
        & \textbf{(mi)} & \textbf{(mi)} & \textbf{Fare (\$)} & \textbf{(\$/mi)} & \textbf{(\$/mi)} \\
        \midrule
        \textbf{DTLA} & 16.6 & 10.0 & 50 & 3.01 & 5.00 \\
        \textbf{LGB}  & 20.3 & 17.0 & 62 & 3.05 & 3.65 \\
        \textbf{WDHL} & 28.2 & 20.0 & 78 & 2.77 & 3.90 \\
        \textbf{UVS}  & 29.7 & 14.0 & 57 & 1.92 & 4.07 \\
        \textbf{ANH}  & 34.6 & 29.0 & 77 & 2.23 & 2.66 \\
        \textbf{HWD}  & 14.6 & 11.0 & 57 & 3.90 & 5.18 \\
        \textbf{PSD}  & 19.3 & 13.0 & 57 & 2.95 & 4.38 \\
        \textbf{BVH}  & 13.6 & 8.5  & 45 & 3.31 & 5.29 \\
        \bottomrule
    \end{tabular}
\end{table}

\begin{table*}[h]
    \centering
    \caption{Composition of operating costs.}
    \label{tab:operating_cost}
    \begin{tabular}{l l l l}
        \toprule
        & \textbf{Cost Unit} & \textbf{Metric from eUAMVRP-NL} & \textbf{Source} \\
        \midrule
        \textbf{Energy cost} & \$0.28 per kWh & Total energy consumption & \cite{bureau_of_labor_statistics_average_2025} \\
        \textbf{Pilot cost} & \$61 per hour  & Total aircraft hours & \cite{economic_research_institute_helicopter_2026} \\
        \textbf{Maintenance cost} & \$0.19 per ASM & Total available seat mile & \cite{joby_joby_2021} \\
        \textbf{Insurance cost} & 3.5\% of the aircraft cost & Fleet size & \cite{rimjha_demand_2020} \\
        \textbf{Battery replacement cost} & \$35{,}000 battery per 1{,}500 flight hours\ & Total aircraft hours & \cite{rimjha_demand_2020} \\
        \textbf{Vertiport operation cost} & \$1{,}000{,}000 per vertiport per year\ & -- & \cite{wisk_aero_positive_2023} \\
        \textbf{Administrative / overhead cost} & 9\% of the total OPEX\ & -- & \cite{federal_aviation_authority_benefit-cost_2018} \\
        \bottomrule
    \end{tabular}
\end{table*}

\subsection{Battery Model and Energy Consumption}

\begin{figure}[h!]
    \centering
    \includegraphics[width=0.7\linewidth]{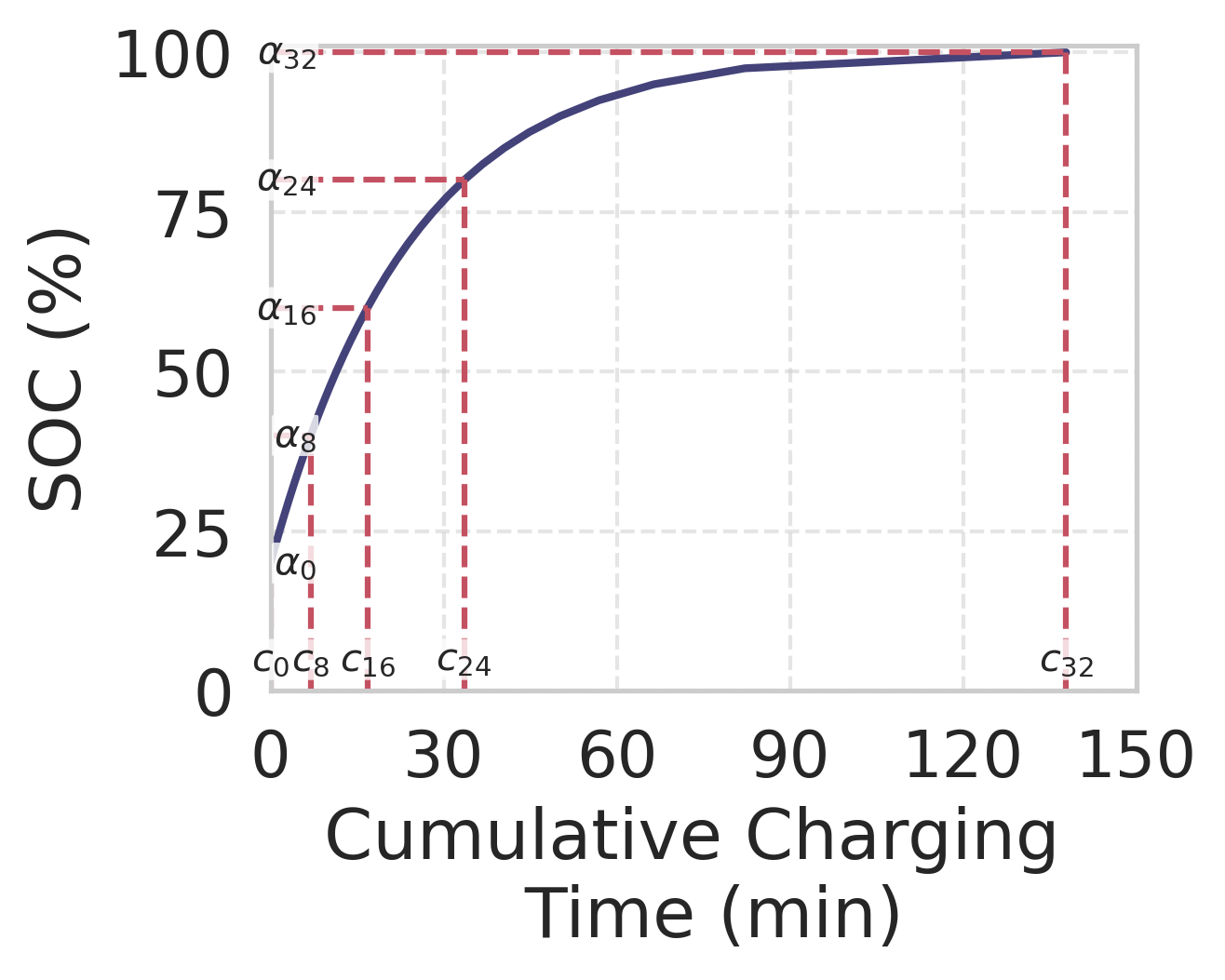}
    \caption{The non-linear battery charging model used in the paper.}
    \label{fig:battery}
\end{figure}

We adopt the battery charging model described in \cite{onat_evaluating_2024}, which uses public experimental
charging data based on the battery pack installed in the Lucid Air Dream Edition, an electric vehicle. The vehicle is equipped with a 118 kWh battery capacity. The battery reaches 80\% SoC from 10\% in only 33 minutes with a 350 kW charger \cite{electric_vehicle_database_electric_2022}. The mathematical relation between charge power and SoC of the battery is described by applying a piecewise linear approximation to the experimental
data. The charge power remains stable when the SoC of the battery is between 0\% and 20\%, and starts to
decrease linearly from 20\% SoC until the charge power reaches 0 kW at 100\% SoC. The charging time is derived by integrating the charge power over time. Figure \ref{fig:battery} visualizes the modeled battery charging time used in eUAMVRP-NL.

\subsection{UAM Capital Cost and Operating Cost}
\label{sec:cost_models}

We consider fleet acquisition cost, real estate cost, and vertiport construction cost to be the capital costs of establishing the UAM network. Joby estimates each eVToL to cost 1.3 million USD \cite{joby_joby_2021}. The real estate cost depends on both land prices and the size of the vertiport. The area of the vertiport depends on the number of operations at the vertiport and the number of parking pads. We use the equation in \cite{onat_urban_2024} to estimate the vertiport footprint, in units of 10 thousand square feet, to be $1.3333 + 0.1667 \times \text{number of pads}$. We use Zhang and Arnott's \cite{zhang_computing_2011} estimate of land prices in the Greater Los Angeles Metro region and convert them into 2025 dollar values using the Producer Price Index (PPI) \cite{st_louis_federal_reserve_producer_2026}. The land values are listed in Tab. \ref{tab:market_values}.

\begin{table}[h]
    \centering
    \caption{Vertiport code and land values.}
    \label{tab:market_values}
    \begin{tabular}{llr}
        \toprule
        \textbf{City Name} & \textbf{Vertiport Code} & \$/$\textbf{ft}^\textbf{2}$ \\
        \midrule
        Los Angeles International Airport & LAX  & 226.14 \\
        Downtown Los Angeles & DTLA & 219.15 \\
        Long Beach & LGB  & 242.18 \\
        Woodlandhills (Van Nuys Airport) & WDHL & 104.97 \\
        Burbank (Universal Studio) & UVS  & 52.77 \\
        Anaheim & ANH  & 88.66 \\
        Hollywood & HWD  & 88.66 \\
        Pacific Palisades & PSD  & 200.00 \\
        Beverly Hills & BVH  & 251.41 \\
        \bottomrule
    \end{tabular}
\end{table}

The construction cost for each vertiport (excluding LAX) is estimated based on "Vertibase" values suggested by Riedel at \$800k USD. \cite{riedel_perspectives_2022} We assume LAX to be a "Vertihub" costing \$7 million USD, given the need for 2 TLOFs at this location. 

The operating costs and the relevant sources are listed in Tab. \ref{tab:operating_cost}. The makeup of the operating costs includes the Energy Cost (consumed in each flight), pilot costs assumed from equivalent costs of a commercial helicopter pilot $\times$ total aircraft hours, maintenance cost (assumed for available seat mile), insurance costs for each aircraft (based on the fleet size), and a battery replacement cost (calculated for every 1500 flight hours). We assume the vertiport operation cost to be fixed at 1 million USD per vertiport per year \cite{wisk_aero_positive_2023}. The administrative overhead is estimated using the Federal Aviation Administration's estimates for legacy airlines \cite{federal_aviation_authority_benefit-cost_2018}.

\subsection{Experiment Setup}

We consider a baseline case where UAM flights are priced statically at \$3 per mile \cite{joby_commercializing_2021}. We consider a fleet size of 20, 30, 40, 50, 60, and 70 UAM aircraft in the airport access market. The optimization models are implemented using the Python API of Gurobi with an Intel Xeon Gold 6326 2.9 GHz CPU.

\section{Results and Discussions}
\label{sec:results}

In this section, we first explore the solution to the joint-supply-demand pricing problem, from which we recover both the UAM fare and the size of the market that UAM captures. We then discuss the various operational metrics we recover from the solution to the eUAMVRP-NL and the overall economic feasibility of UAM in the LAX airport access market.

\begin{figure}[h!]
    \centering
    \includegraphics[width=1.0\linewidth]{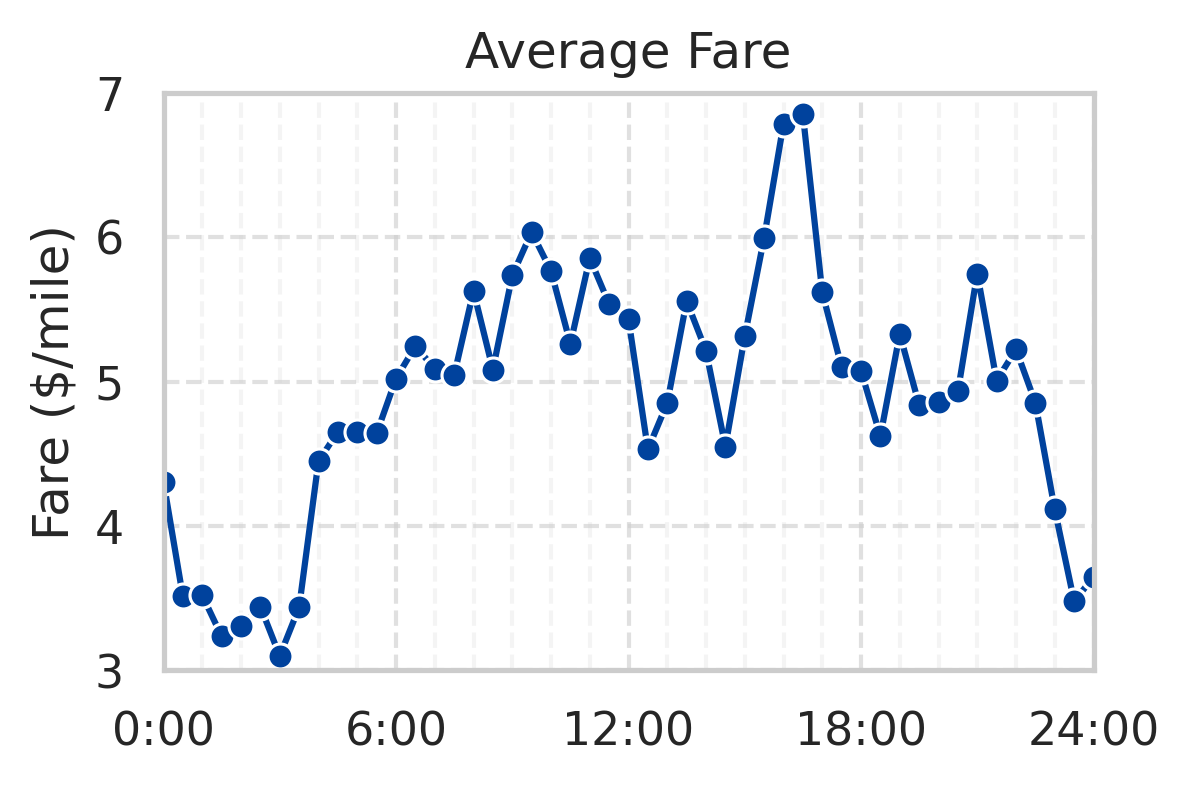}
    \caption{Average UAM fare in RASM over time given a fleet of 30 aircraft.}
    \label{fig:average_rasm}
\end{figure}

\begin{figure*}[h]
    \centering
    \includegraphics[width=1.05\linewidth]{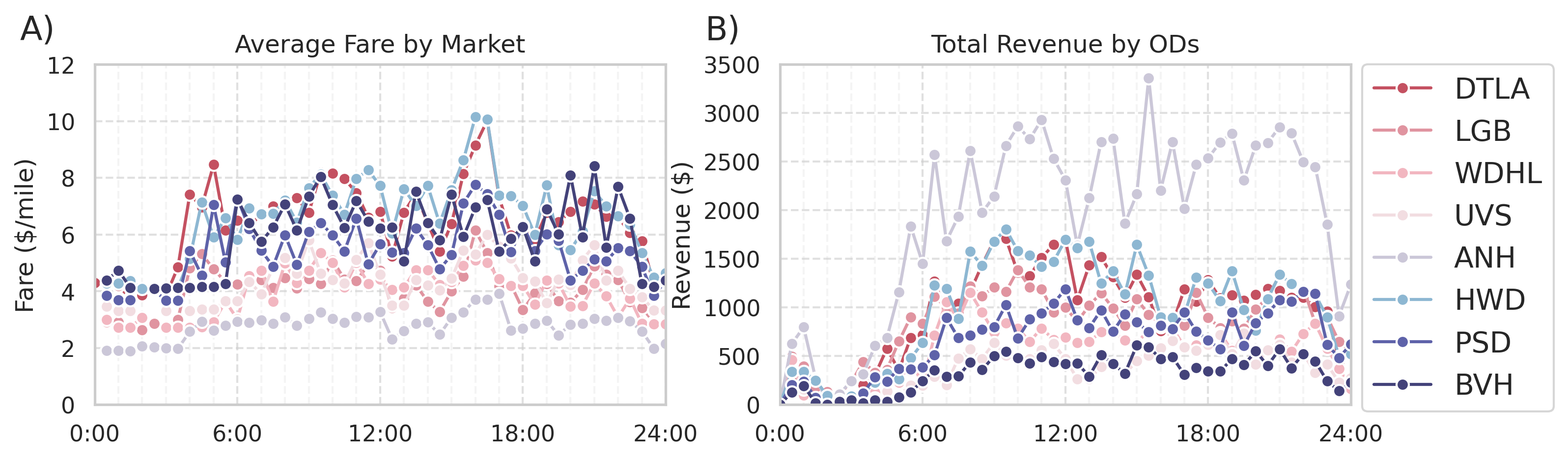}
    \caption{Average UAM fare and total revenue by OD markets given a fleet of 30 aircraft.}
    \label{fig:rasm_by_od}
\end{figure*}

\subsection{Pricing}

Figure \ref{fig:average_rasm} shows the average fare charged per passenger-mile over the course of the day. We observe a strong correlation between the travel time advantage of UAM flights, as indicated in Fig. \ref{fig:tnc_trip_time}-\ref{fig:uam_trip_time}, and the fare determined by the solution to the pricing problem. The average fare can be as much as twice the lowest fare during the peak hour of the day - a significant change in prices. The results reveal that in the LAX airport access market, UAM operators have strong pricing power and capitalize on rush hours.

It is also worth pointing out that the pricing results support our previous observation that the travel time advantage of UAM is vulnerable. In the early mornings, UAM flights are priced at a fare between \$3 and \$3.5, which is, in fact, lower than the average Uber fare after accounting for the difference between ground travel distance and air travel distance. UAM is priced lower than Uber during these times because the transition time at vertiports and the last and first mile trip times drastically increase the total UAM trip time, decreasing the utility function value of UAM. Ground travel without congestion is highly efficient.

Figure \ref{fig:rasm_by_od}A reveals that the pricing of UAM flights depends not only on the time of day but also on OD markets. Whereas the Hollywood (HWD) and Downtown Los Angeles (DTLA) market is priced at a fare of up to \$10 per mile in the afternoon, the Anaheim (ANH) market is priced at a fare between \$3 and \$4 per mile. Both are large revenue-generating markets, as indicated in Fig. \ref{fig:rasm_by_od}B but they exhibit drastically different pricing policies. The  Hollywood market, shown in Tab. \ref{tab:uber}, has the highest RpGM and the second highest RpAM in TNC services. Given that it is already an expensive market, UAM flights are also priced higher. The Downtown Los Angeles market has a relatively low RpGM but one of the highest RpAMs. This large gap highlights the circuitous nature of the fastest route on the ground. Therefore, even though TNCs such as Uber do not charge a hefty price in the Downtown Los Angeles market from the perspective of the distance traveled, it can be one of the most profitable markets for UAM due to the gap between ground and air distance. On the other hand, the Anaheim market has the lowest RASM but is the highest revenue generating market for UAM. It is the farthest market from LAX in the network by both ground and air distances. The speed of UAM aircraft is much higher in cruise than in transition, takeoff, and landing. Even though Anaheim has a low fare, it allows for the efficient use of the fleet. The revenue per hour each aircraft can generate increases significantly in the Anaheim market. This is an observation we will explore further in the following sections.

\subsection{Economics of UAM Operations }
\label{sec:financial_projection}

We compute the various cost items explained in Sec. \ref{sec:cost_models} using the solution to eUAMVRP-NL which provides detailed aircraft schedules that include revenue flights, repositioning flights, and charge schedules. We can also recover the load factor of each flight and the total revenue generated by each aircraft in the fleet. Table \ref{tab:finance} compares the operational and financial metrics between the fixed pricing scheme of \$3/mi and the variable pricing policy output by the joint-supply-demand pricing model explained in the previous section.

\begin{table}[h]
    \centering
    \caption{System utilization, cost structure, and profitability comparison between the fixed pricing of \$3/mi and variable pricing for a fleet of 30 aircraft.}
    \label{tab:finance}
    \begin{tabular}{lcc}
        \toprule
        Pricing Scheme & \$3/mi & Variable Pricing\\
        \midrule
        Fleet Size & 30 & 30 \\
        Total number of TLOFs & 10 &  10 \\
        Total number of pads & 57    &  59\\
        \textbf{Total CapEx (\$) } & \textbf{88.4M} & \textbf{89.4M} \\
        Daily number of revenue flights & 1,071 & 936 \\
        Daily number of repositioning flights & 128 & 106\\
        Percentage of repositioning flights (\%) & 10.68 & 10.17\\ 
        Number of revenue passengers & 3,837 & 2,797 \\
        Load factor (\%) & 80.0\% & 67.1\% \\
        Number of aircraft hours (hrs) & 402 & 371 \\
        Revenue per aircraft hour (\$) & 344 & 560 \\
        Revenue aircraft miles (mi) &  12,987   & 15,181   \\
        Total aircraft miles (mi) &   14,387    &  16,648   \\
        \textbf{RASM (\$)} & \textbf{2.402} & \textbf{3.123} \\ 
        Energy cost per ASM  (\$) &  0.095      &    0.078  \\
        Pilot cost per ASM   (\$) & 0.426    &    0.340  \\
        Battery replacement cost per ASM &    0.163   &  0.130    \\
        Maintenance cost per ASM (\$)&    0.190    & 0.190    \\
        Insurance cost per ASM (\$) &      0.065  &   0.056     \\
        Vertiport operation cost per ASM (\$)&   0.428     &  0.370   \\
        Administrative cost per ASM (\$)&   0.135       &   0.115     \\
        \textbf{CASM (\$)}  & \textbf{1.503}      &     \textbf{1.279}       \\

        Daily revenue (\$) & 138,203 & 207,938 \\
        Daily operating cost (\$)& 86,486 & 85,169 \\
        \textbf{Daily profit (\$)} & \textbf{51,717} & \textbf{122,769} \\
        \bottomrule
    \end{tabular}
\end{table}

\begin{figure}[h!]
    \centering
    \includegraphics[width=1.0\linewidth]{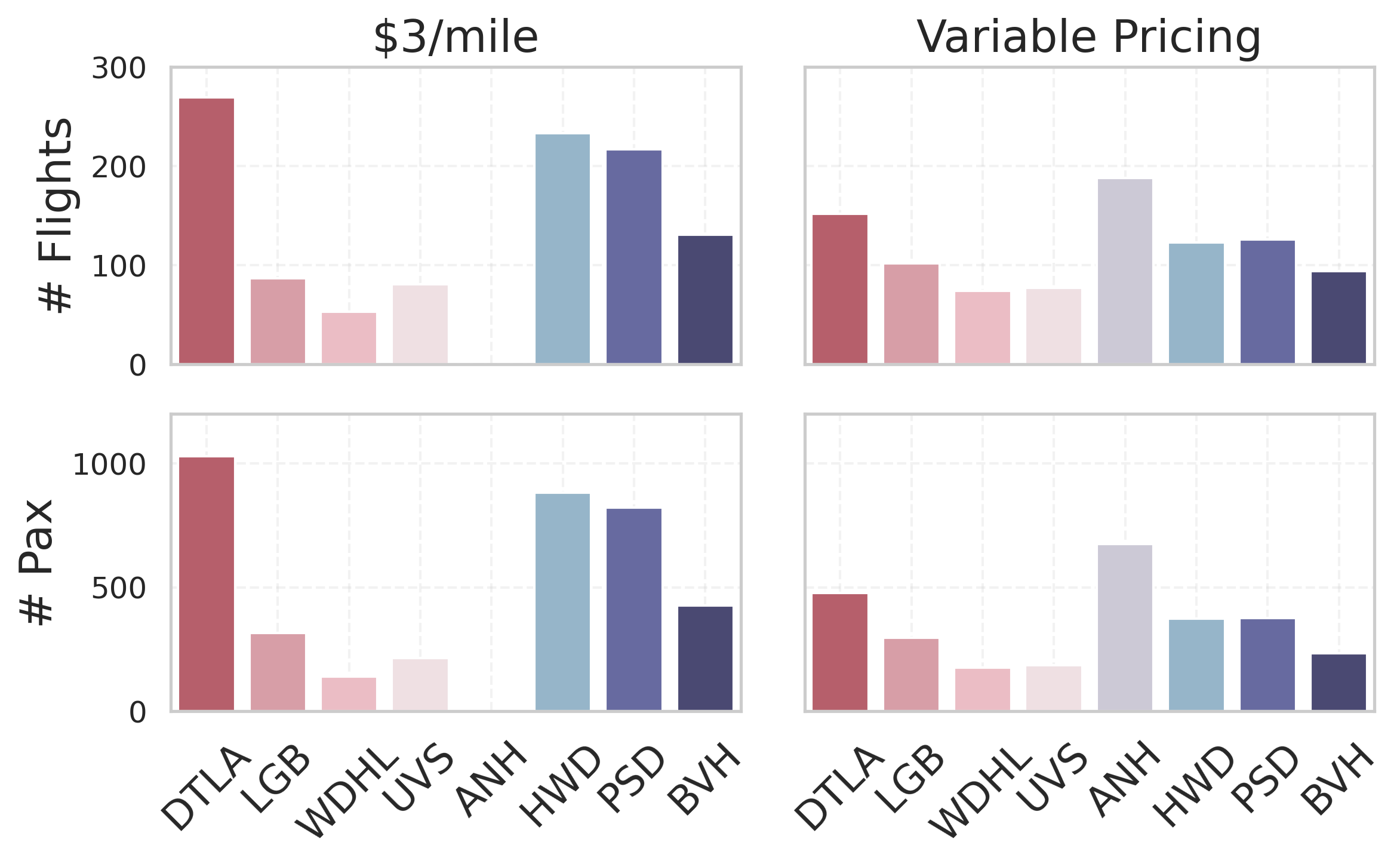}
    \caption{The number of passengers and revenue flights served under fixed and variable pricing schemes.}
    \label{fig:market}
\end{figure}

We see a 137.4\% increase in profits, from \$51,717 to \$122,769 per day. The higher profits are driven by both a 50\% increase in revenue and a decrease in operating costs. We see a 62\% increase in revenue per aircraft-hour, while RASM only has a 30\% increase. This suggests that the higher revenue is closely related to a more efficient use of the fleet rather than the increase in the average fare charged. The number of revenue passengers served decreases from 3,837 to 2,797, and the number of revenue flights decreases from 1,071 to 936, while the number of total aircraft miles increases by 15.7\%. This suggests that aircraft are serving longer OD markets under variable pricing. Aircraft spend less time transitioning in and out of the vertiport but more time in cruise. Therefore, total distance traveled is greater under the variable pricing scheme. This observation is further supported by evidence in the market shift shown in Fig. \ref{fig:market}. The Anaheim market, in which UAM was previously dominated by TNC in the utility model under the \$3 pricing scheme, has now become the largest market by both the number of revenue passengers and the number of revenue flights. As previously discussed, Anaheim is the farthest market from LAX. The increase in total aircraft miles and revenue aircraft miles is driven by the growth in the Anaheim market. 

The shift to longer OD markets also helps to cut operating costs. We see a 15\% decrease in Cost per Available Seat Mile (CASM). Aircraft consume less energy for the same distance flown in cruise, and the speed of the aircraft is significantly higher during cruise. This not only helps lower the amount of energy consumed per unit distance but also reduces pilot costs per unit distance, both of which experience a clear decrease under the variable pricing scheme. The results suggest that the economies of stage length, that is, the lowering of operating costs when the average distance of revenue flights increases, applies to UAM operations. 

We find that revenue per aircraft hour is a better metric for understanding the performance of UAM operations and the overall profitability than RASM, which is the conventional metric that legacy airlines consider. The 50\% increase in revenue is closer to the 62\% increase in revenue per aircraft rather than the 30\% increase in RASM. The metric revenue per aircraft hour better reflects the operational paradigm of the UAM fleet, as each aircraft completes multiple revenue flights per day and spends a significant amount of time in charge and transitioning in and out of the vertiports. 

The RASM we report under the variable pricing scheme, \$3.123, is similar to the \$3.00 per mile fare claimed by Joby \cite{joby_commercializing_2021}. However, questions remain as to whether such a price is sustainable under the condition that not every revenue flight can achieve a 100\% load factor and that repositioning flights are required.

This comparison between fixed pricing and variable pricing suggests that UAM operators are more likely to behave similarly to TNCs rather than to existing transit operators. The profitability of the UAM service significantly increases under a variable pricing scheme. The results also show that UAM operators benefit from having complete control over the scheduling of the fleet, as variable pricing and fleet scheduling improve the utilization of the aircraft.

\subsection{Fleet Sizing}

\begin{figure*}[h!]
    \centering
    \includegraphics[width=1.0\linewidth]{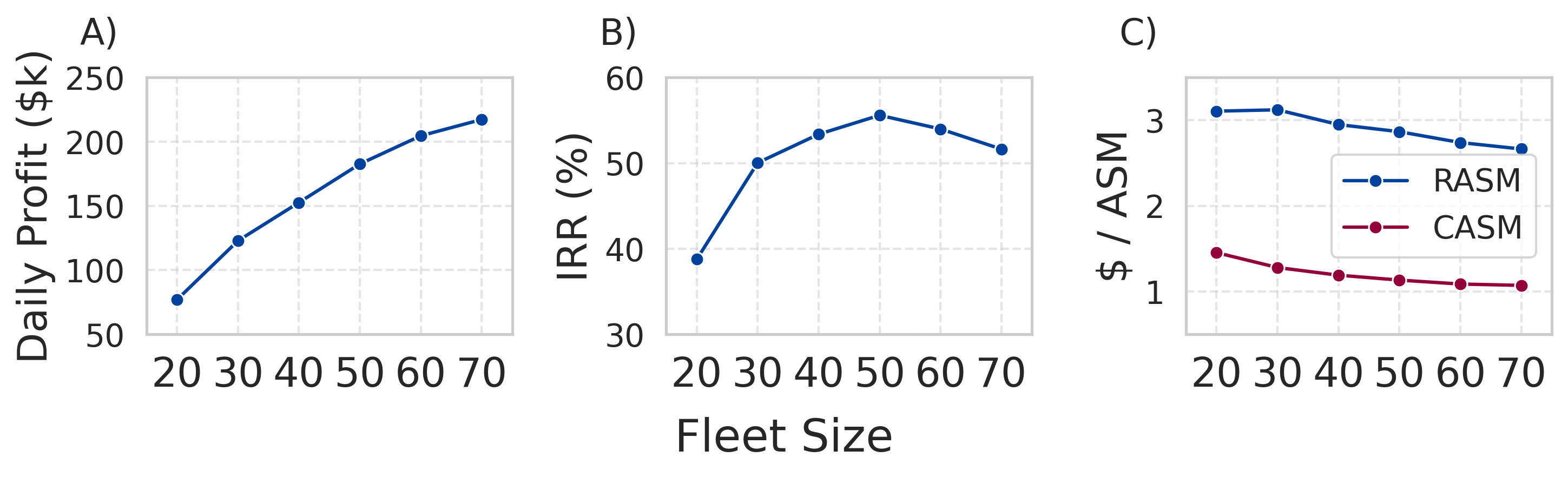}
    \caption{Effect of fleet size on daily operating profit, IRR, CASM, and RASM.}
    \label{fig:return}
\end{figure*}

We are also interested in understanding the impact of different fleet sizes on the LAX airport access market. We compute the Internal Rate of Return (IRR) using the following formula:

\begin{equation}
    0 = \sum_{t=1}^{15} \frac{\text{Operating Profit per Year}}{(1+IRR)^{t}} - \text{Capital Cost}
\end{equation}

Figure \ref{fig:return}B shows that the IRR peaks at 50 aircraft in the market we consider. Although a larger fleet leads to an increase in operating profit, shown in Fig. \ref{fig:return}A, it also leads to an increase in fleet and land acquisition costs, as larger vertiports are required to accommodate the larger fleet. We also notice a marginally decreasing return in profits as fleet size grows. Doubling the fleet from 20 to 40 aircraft leads to a \$75,000 increase in daily profits, but doubling the aircraft from 30 to 60 only leads to a \$60,000 increase in daily profits.

Figure \ref{fig:return}C shows that an increase in fleet size leads to a decrease in both RASM and CASM. The most profitable segment of the market becomes saturated at a fleet size of 30 aircraft. To capture more passengers for UAM flights, UAM operators need to lower the fare, which leads to a gradual decrease in RASM. CASM also gradually decreases as fleet size increases. A larger fleet can achieve economies of scale with respect to vertiport operation costs, which are shared among all aircraft in the fleet. However, the rate at which RASM decreases outpaces the rate at which CASM decreases, as a more significant portion of the operating costs is related to the volume of flights.

\begin{figure}[h!]
    \centering
    \includegraphics[width=1.0\linewidth]{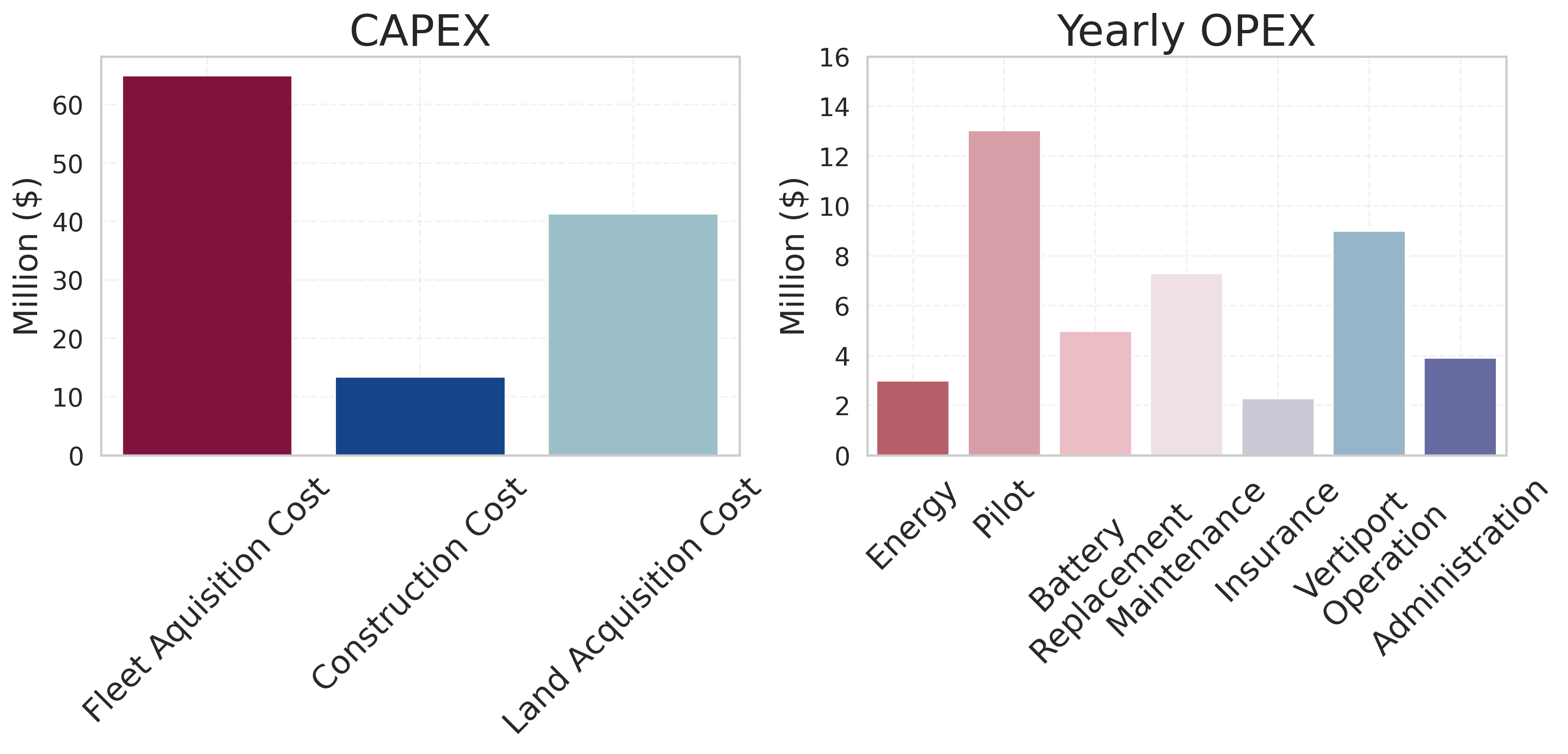}
    \caption{Capital and operating cost of 50 aircraft in the LAX airport access market.}
    \label{fig:50}
\end{figure}

\subsection{Financial Projection}

We compute the capital cost for operating a fleet of 50 aircraft, which has the highest IRR, and the yearly operating cost. In Fig.\ref{fig:50}, we find that the cost of labor, namely pilots, is the single largest component, followed by vertiport operating costs and maintenance. On the other hand, energy and battery replacement costs constitute a smaller percentage of the total operating cost. This suggests that electric motors and batteries can be a promising powertrain for UAM considering the operating cost.

\begin{figure}[h!]
    \centering
    \includegraphics[width=1.05\linewidth]{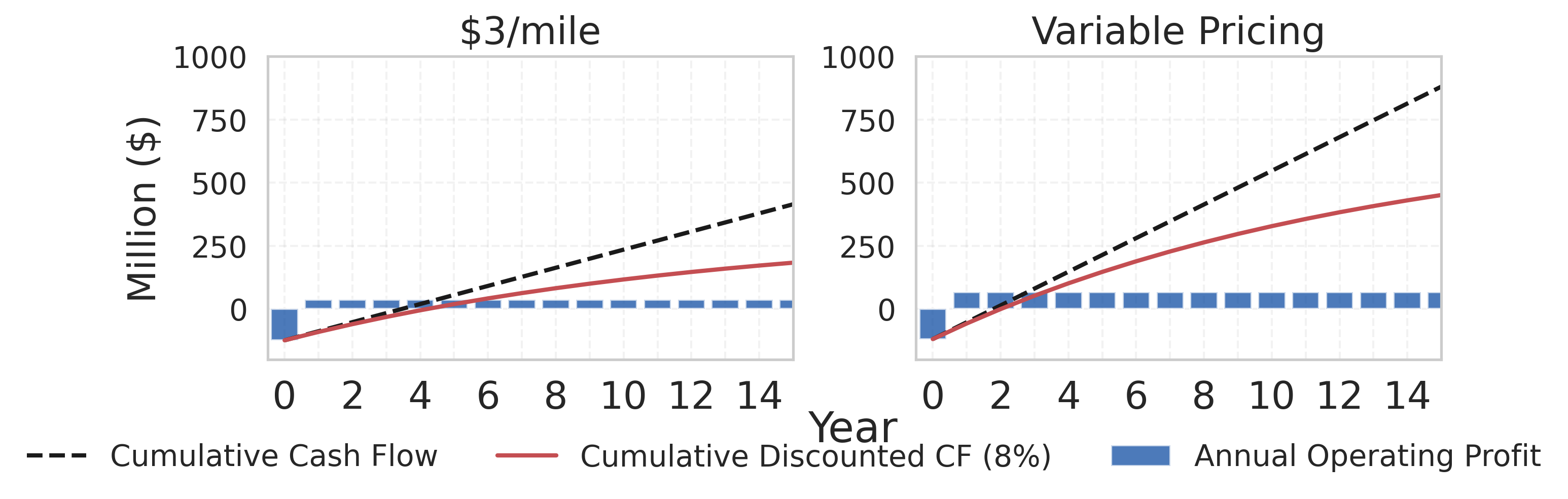}
    \caption{Financial forecast for the UAM service given a fleet of 50 aircraft.}
    \label{fig:forecast}
\end{figure}

Shown in Fig. \ref{fig:forecast}, we find that variable pricing can cut the discounted payback period of the capital cost from 5 to 3 years. Again, this suggests that UAM in the airport access market is more similar to a TNC service than to a transit service. The variable pricing scheme provides drastically better financial incentives for UAM operators. 

\section{Conclusion}

In this paper, we propose a two-step framework for optimizing profitability and estimating the economic feasibility of a UAM service. We leverage a joint-supply-demand optimization formulation to compute the profit-maximizing pricing strategy for UAM operators and use eUAMVRP-NL to solve the fleet scheduling problem, from which the solutions we obtain are used to estimate the capital cost, operating revenue, and operating cost. We conduct a case study for the airport access market centered on LAX. We discover that variable pricing can drastically increase the profitability of the UAM service and that UAM operators benefit from adopting TNC-like fleet management and pricing strategies. We find that the airport access market can be highly profitable, given sufficient travel time advantages in UAM and affordable land acquisition or rental costs.

For future endeavors, we seek to model the competition between TNCs and UAM service providers through a game theory approach. Currently, we assume TNC's pricing policy is static, not only in the sense that the fare is fixed throughout the day, but also in the idea that TNCs will not change their pricing policies in response to UAM's entrance into the market. The response of TNCs can be reflected in their pricing, which has a direct impact on the UAM market share. Understanding the interaction between existing service providers and new UAM service operators can provide more accurate insights into the operational paradigm and pricing policy of a UAM service.

\bibliographystyle{IEEEtran}
\bibliography{references}

\vspace{12pt}

\end{document}